\documentclass[aps,preprint]{revtex4}
\usepackage{amsmath}
\usepackage{epsfig}
\usepackage{float}
\usepackage{dcolumn} 
\usepackage{physics}

\DeclareMathAlphabet{\mathpzc}{OT1}{pzc}{m}{it}
\newcommand{\be}{\begin{equation}}
\newcommand{\ee}{\end{equation}}
\newcommand{\bea}{\begin{eqnarray}}
\newcommand{\eea}{\end{eqnarray}}
\newcommand{\beas}{\begin{eqnarray*}}
\newcommand{\eeas}{\end{eqnarray*}}

\begin{document}
\title{\bf Dirichlet spectra of the paradigm model of complex PT-symmetric potential: $V(x)=-(ix)^N$}  
\author{$^1$Zafar Ahmed, $^2$Sachin Kumar, $^3$Dhruv Sharma\\
$^1$Nuclear Physics Division, $^2$Theoretical Physics  Section\\ Bhabha Atomic Research Centre, Trombay, Mumbai, 400085, India \\
$^3$Department of Physics, National Institute of Technology, Rourkela, 769008, India}
\email{1:zahmed@barc.gov.in, 2:sachinv@barc.gov.in, 3:sharmadhru@gmail.com}
\date{\today}
\begin{abstract}
\noindent
So far the spectra $E_n(N)$ of the paradigm model of complex PT(Parity-Time)-symmetric potential $V_{BB}(x,N)=-(ix)^N$ is known to be analytically continued for $N > 4$. Consequently, the well known eigenvalues of the Hermitian cases ($N=6,10$) cannot be recovered. Here, we illustrate Kato's theorem that even if a Hamiltonian $H(\lambda)$ is an analytic function of a real parameter $\lambda$, its eigenvalues $E_n(\lambda)$ may not be
analytic at finite number of Isolated Points (IPs). In this light, we present the Dirichlet spectra $E_n(N)$ of $V_{BB}(x,N)$ for $2\le N<12$ using  the numerical integration of Schr{\"o}dinger equation with $\psi(x=\pm \infty)=0$ and the diagonalization of $H=p^2/2\mu+V_{BB}(x,N)$ in the harmonic oscillator basis. We show that these real discrete spectra are consistent with the most simple two-turning point CWKB (C refers to complex turning points) method provided we choose the maximal turning points (MxTP) [$-a+ib,a+ib, a, b \in {\cal R}$] such that $|a|$ is the largest for a given energy among all (multiple) turning points. We find that  $E_n(N)$ are continuous function of $N$ but non-analytic (their first derivative is discontinuous) at IPs $N=4,8$; where the Dirichlet spectrum is null (as $V_{BB}$ becomes a Hermitian flat-top potential barrier). At $N=6$ and $10$, $V_{BB}(x,N)$ becomes a Hermitian well and we recover its well known eigenvalues.
\end{abstract}
\maketitle
\section{Introduction}
A non-Hermitian (complex) Hamiltonian $H$ is PT(Parity-Time)-symmetric [1], if it is invariant
under the joint action of Parity (P: $x\rightarrow -x$) and Time-reversal
(T: $i\rightarrow -i$) transformations. Under the  PT-symmetry of a Hamiltonian there are two parametric regimes of  unbroken (exact) and broken PT-symmetry as per the behavior of eigenstates of $H$. If    eigenstates of $H$ are also eigenstates of PT: PT$\psi_n=(-1)^n \psi_n$ [1] PT-symmetry is called unbroken (exact) and eigenvalues are real. This happens below or above a critical value of a potential parameter. For instance, for $V_{BB}(x,N)=-(ix)^N$ the entire spectrum is real and PT-symmetry is exact if $N\ge 2$. Otherwise, the PT-symmetry is spontaneously broken and eigenvalues are complex conjugate pairs (see Fig. 4, for $N<2$). The corresponding eigenstates flip under PT: PT$\psi_{E}=\psi_{E^*}$. In the parametric domain
of broken PT-symmetry all or most of the eigenvalues are complex conjugate pairs excepting a few low lying ones. But in the  unbroken domain (e.g., $N\ge 2$) all eigenvalues are real.

The first Complex PT-Symmetric Potential(CPTSP) $V_{BB}(x,N)$ proposed by Bender and Boettcher [1] has brought a paradigm shift in quantum mechanics. It proposes that even non-Hermitian Hamiltonians can have real discrete spectrum. Based on the numerical computations they conjectured that the entire discrete spectra of $V_{BB}(x, N)$ for $N\ge 2$ were real. Using  spectral determinants, the Bethe ansatz, the Baxter relation,the monodromy group, and a broad spectrum of techniques used in conformal quantum field theory, Dorey et al [2] proved that the spectrum $V_{BB}(x,N)$ for $N\ge 2$ is entirely real, positive and discrete [3].

The parametric evolution of the spectra $E_n(N)$ (1) [1]
of $V_{BB}$ is  analytically continued for $N > 4$, ignoring the fact that for $N>4$ there are more than one pair of complex turning points in contrast to the cases when $2\le N\le 4$. Here, we illustrate Kato's [4] theorem that a Hamiltonian $H(\lambda)$ which is an analytic function of a real parameter $\lambda$, its eigenvalues ($E_n$) need not be  analytic function of $\lambda$, instead they may be non-analytic at  Isolated Points (IPs). Employing three methods, we present the parametric evolution of the Dirichlet spectra for the first five eigenvalues $E_n(2 \le N <12)$, which are continuous but non-analytic at $N=4,8$. At these two IPs the first derivative is discontinuous and most distinctly $V_{BB}$ becomes a flat-top Hermitian barrier, it is where the Dirichlet spectrum is null and one gets discrete reflectivity zeros [5,6].

Conventionally, in quantum mechanics, there could be three kinds of states characterizing discrete (quantized) spectra: (i) bound states, (ii) perfect transmission (zero reflectivity) states and (iii) complex energy resonant states. For obtaining the bound state spectrum of a one dimensional potential well, one imposes the Dirichlet boundary condition on the wave function i.e. $\psi(\pm \infty)=0$. Giving up this common practice, in Ref. [1] an uncommon and elegant method of eigenvalue problem on complex contours has been adopted. A notion of wedges has been used, wherein the Schr{\"o}dinger equation is solved numerically along the anti-stokes lines and WKB solutions were matched at an asymptotic distance. It has been found that a simple 2-turning point Complex WKB (CWKB) formula
\begin{equation}
E_n^{BB}(N) = \left[ \frac{\Gamma(3/2+1/N) \sqrt{\pi}(n+1/2)}{\sin (\pi/N)\Gamma(1+1/N)}\right ]^{\frac{2N}{N+2}},~ n=0,1,2,.., \quad N\ge 2
\end{equation}
using complex turning points as [1]
\begin{equation}
x_{-}(N)=E^{1/N} \exp[i\pi(3/2-1/N)], \quad x_+(N) =E^{1/N} \exp[-i\pi(1/2-1/N)].
\end{equation}
reproduces their numerically obtained spectra well  for $N \ge 2$, where PT-symmetry is exact. More importantly, Eq.(1) also represents the analytic continuation of $E_n(N)$ of $V_{BB}$ for  $N > 4$ as suggested in [1].

The spectra of the potential $V_{BBM}(x)=x^{2M}(ix)^\epsilon$ for $-M \le \epsilon \le 0$  has also been studied [7] for $M>1$. However, the variation of the parameter $\epsilon$ is limited in $(-M,0]$ after which eigenvalues are analytically continued. In a more lucid explanation of the methodology of the wedges in complex plane for the spectra of $V(x)=x^2(ix)^{\epsilon}(N=\epsilon +2)$, a semi-classical expression for $E_n(\epsilon)$ similar to (1) and (8) (see below) appears. Where, in place of $\sin((2k+1) \pi/N)$ there occurs a curious factor of $\cos(\gamma)$. Instead of an explicit expression for $\gamma$,  a detailed prescription for only integral values of $\epsilon$ has been given [8]. On one hand, the existence of independent families (even for $V(x)=x^{6}$) of real spectra for each $N$ ($n$ fixed) has been  professed [8] and called PT-symmetric spectra. While, on the other hand, analytically continued evolution of  $E_n(N)$ for $V_{BB}(x,N)$  as found in [1], has been studied and confirmed by the method of  wedges in complex $x$-plane by other authors in the parametric domain of $1< N \le 5$ in various ways [9-12].

Eventually, their [1,3-7,9-12] eigenstates  $\psi_n(x)$ do not essentially vanish asymptotically on the real line. However, the elegant method of wedges in the complex $x$-plane [1] unifies  two disparate situations to produce real discrete spectra of the bound states and for the Hermitian flat-top barrier $V_{BB}(x,4)=-x^4$, it has produced  (above the barrier) discrete zeros of reflectivity $(R(E_n)=0)$ [5,6]. The present work can be seen as an attempt to find a unique $E_n(N)$ ($n$ fixed) which will be non-analytic at IPs and this can be understood from  the discussion given below in the next section.
\begin{figure}[t]
\centering
\includegraphics[width=7 cm,height=5cm]{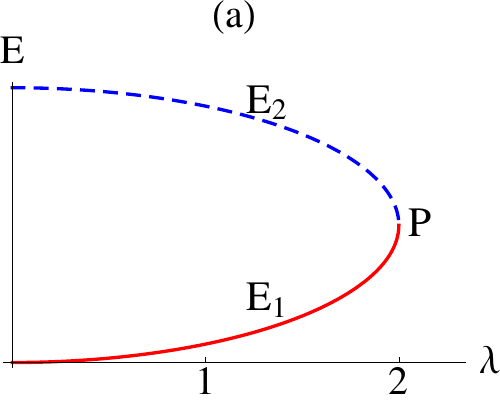}\hspace {.5 cm}
\includegraphics[width=7 cm,height=5cm]{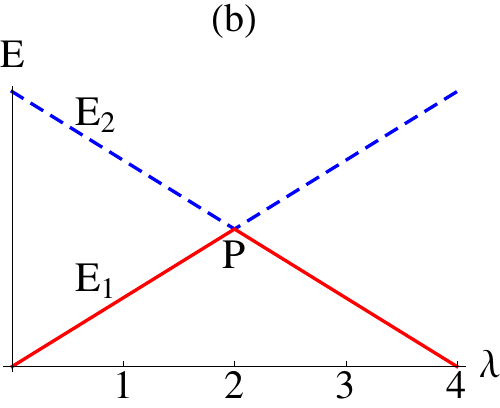}\vspace{.5 cm}\\
\includegraphics[width=7 cm,height=5cm]{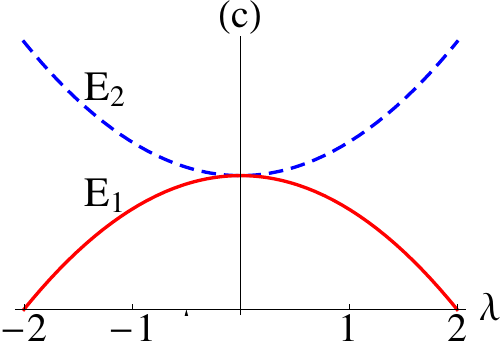}\hspace{.5 cm}
\includegraphics[width=7 cm,height=5cm]{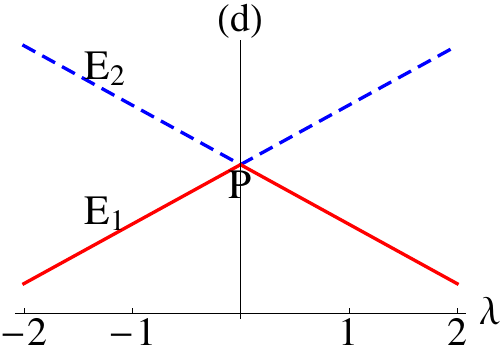}
\caption{Scenarios of non-analyticity in the the parametric evolution of eigenvalues of the analytic non-Hermitian $A(\lambda), D(\lambda)$ and Hermitian  $ B(\lambda)$ Hamiltonians (3). The correct evolution of two eigenvalues is represented by $E_1$ (solid lines) and $E_2$ (dashed lines), where the point P is the point of non-analyticity. In the case (a) the point P  is called Exceptional Point where two real eigenvalues coalesce and $\left .\frac{dE}{d\lambda}\right|_{\lambda=2}= \infty$. In cases (b,d), the point P is IP where the derivative is discontinuous. The evolution of $E(\lambda)$ in (c) is analytic suggesting that in cases (b,d)  the evolution of one  eigenvalue cannot comprise in both dashed and solid lines.} 
\end{figure}

\section{Non-analyticity of $E_n(\lambda)$ at an IP for analytic $H(\lambda)$}
Usually, if a Hamiltonian is an analytic function of a real parameter $\lambda$, the parametric evolution of eigenvalues $E_n(\lambda)$ is also an analytic
function of $\lambda$. By analyticity  one means that $E_n(\lambda)$ is  differentiable (left and right derivatives equal and finite) at each and every point $(\lambda \in D)$. This is how the Hellmann-Feynman theorem (HFT) [13] for a Hermitian Hamiltonian $\frac{dE_n}{d\lambda}=<\psi_n|\frac{\partial H}{\partial\lambda}|\psi_n>$ holds true. Normally, both $H(\lambda)$ and the eigenvalues $E_n(\lambda)$ are analytic functions of $\lambda$. One simple exceptional case is of the harmonic oscillator $V(x,\omega)=\frac{1}{2}m(\omega^2-\omega_0^2)x^2$ which is analytic function of $\omega$ but its eigenvalues  $E_n=(n+1/2)\hbar \sqrt{\omega^2-\omega_0^2}$  are non-analytic at $\omega=\omega_0$, $dE_n/d\omega=\infty$ and $\omega=\omega_0$ is the exceptional point of $V(x,\omega)$. It is rightly so because for $\omega \le \omega_0$ this potential well gets inverted to become a barrier which is devoid real discrete spectrum. In the following, we give example of three  $2 \times 2$ matrices $A(\lambda), B(\lambda)$, and $D(\lambda)$ whose elements are analytic functions of $\lambda$ but their respective eigenvalues $E^A_{1,2}, E^B_{1,2}$ and $E^D_{1,2}$ are not so. 
\begin{small}
\begin{equation}
A(\lambda)= \left(\begin{array}{cc}  5 & 2- \lambda \\ 2 + \lambda & 5 \end{array} \right ),
B(\lambda)=\left(\begin{array}{cc} 5 & \lambda-2\\ \lambda -2 & 5 \end{array}\right), C(\lambda) = \left(\begin{array}{cc} 5 & \lambda^2 \\
\lambda^2 & 5 \end{array} \right), 
D(\lambda)= \left(\begin{array}{cc}  5 & 4\lambda  \\ \lambda & 5 \end{array} \right ).
\end{equation}
\end{small}
Here the eigenvalue functions corresponding to  these four matrices are:$E^A _{1,2}(\lambda)=5\mp\sqrt{4-\lambda^2}$, $E^B_{1,2}(\lambda)= 5 \mp|\lambda-2|$, $E^C_{1,2}= 5 \mp \lambda ^2$,  $E^D_{1,2}(\lambda)= 5 \mp 2|\lambda|$ (see Fig. 1). The matrix $C(\lambda)$
is an ordinary example where both the Hamiltonian and the eigenvalues are analytic in $\lambda$.

The derivative of eigenvalues $E^A_{1,2}(\lambda)$ (becomes $\infty$) does not exist at $\lambda=2$, such a parametric point is called Exceptional Point (EP) [4] of a non-Hermitian Hamiltonian. The spectrum of $V_{BB}$ for $1<N<2$ [1] contains such EPs (see Fig. 4). For such EPs in complex PT-symmetric potentials also see [14,15].

 It can be seen that the eigenvalues $E^B_{1,2},E^D_{1,2}$ have their derivative as discontinuous at an IP (see Fig.1), irrespective of whether the Hamiltonian is Hermitian or non-Hermitian. Despite the discontinuity of the derivatives of eigenvalues, one can readily verify that HFT is satisfied for the Hermitian Hamiltonian $B(\lambda)$ in (3). 

The exclusion/neglect of the IPs in a spectrum will lead one to follow an incorrect evolution of eigenvalues from dashed to solid and vice versa (see Fig.1). Whereas, their inclusion leads to a correct evolution of eigenvalues as $E_1(\lambda)$ (solid lines) and $E_2(\lambda)$ (dashed lines), separately. This is just like one will do in case of the analytic eigenvalues of $E^C_1,E^C_2$ for $C(\lambda)$ (see Fig. 1). In fact these toy models of Hamiltonians (B,D) (3) are the simple illustrations of a very interesting theorem by Kato [2]; it asserts that even for analytic Hamiltonians $H(\lambda)$, the evolution of $E_n(\lambda)$ may have derivative discontinuous at a finite number of IPs. Consequently, the evolution of the spectra will be composed of several piece-wise continuous functions joint continuously at IPs [4].

The choice [1,3-7,9-12] of analytically continuing the eigenvalues, though mathematically elegant, cannot be a necessary physical condition on the parametric evolution of a spectrum of non-Hermitian potential. For instance, for $V_{BB}(x,N)$ as $N$ varies continuously, the shape of the potential changes dramatically for $N = 4,8$. At these values they are Hermitian flat top barriers possessing reflectivity zero [5,6]. We propose that these values are the IPs of $V_{BB}(x,N)$. Next, for $N = 6$ and $N=10$, we have Hermitian potentials but, now these are potential wells, with well known eigenvalues [17]. Reproducing these eigenvalues in $E_{n}(N)$ is most desirable.

\section{The Maximal Turning point} 
On par with the {\it fundamental theorem of algebra}, the classical turning points of a CPTSP namely the roots of $V_{PT}(x)=E$  have been argued [16] to be of the types $(-z^*,z): (-a+ib, a+ib), (-a,a), ic, id,..$; here  $ a,b,c,d \in {\cal R}$. It has been found that the phase space $(x, p)$ is segregated in two parts $(x, p_r)$ and $(x, p_i)$: real and imaginary respectively. In the former, phase-space orbits are symmetric, enclosing a finite area. Whereas, the latter are anti-symmetric, enclosing null area $\--$ justifying the reality of eigenvalues. See these two segregated phase-spaces and orbits in Fig. 2 for $V_{BB}(x,3)=ix^3$ for the first three real discrete eigenvalues (see Table I). 

\begin{figure}[t]
	\centering
	\includegraphics[width=8 cm,height=8.cm]{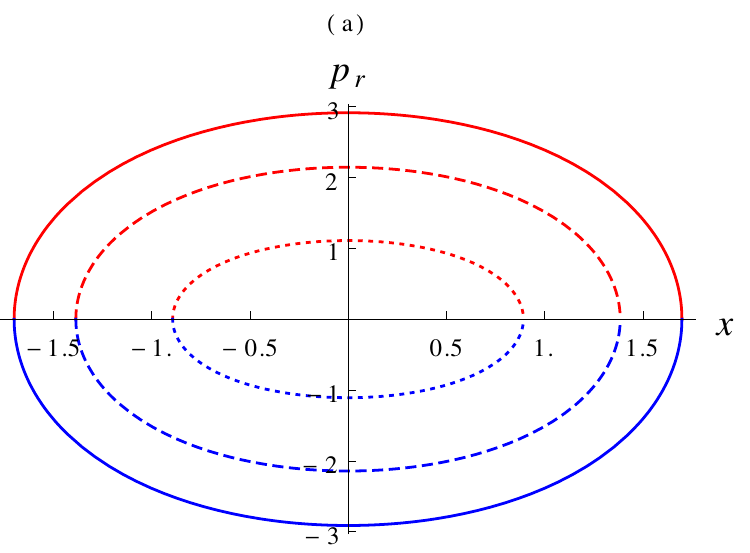}
	\includegraphics[width=8 cm,height=8.cm]{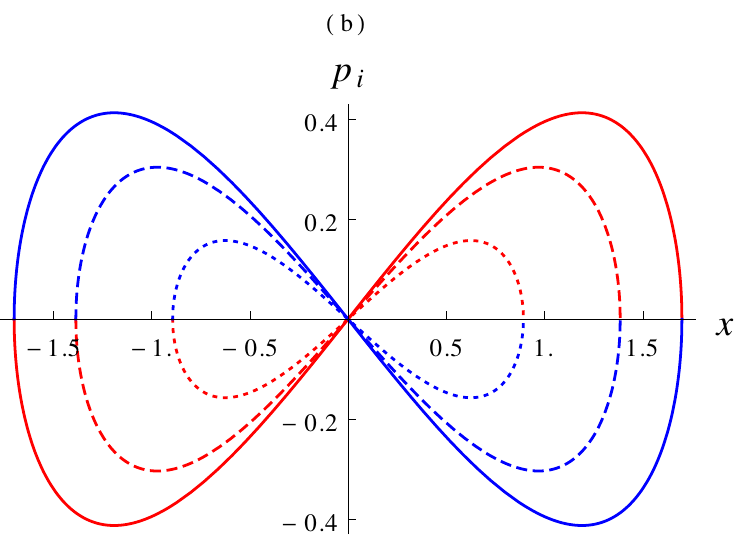}
	\caption{The phase space orbits for $V_{BB}(x,3)=ix^3$ corresponding to first three eigenvalues (see Table I).  (a): in
		the real part of phase space, the orbits are closed enclosing finite area giving rise to real discrete spectra, like in a Hermitian case (b) in the imaginary part of the phase space, the orbits are anti-symmetric enclosing area as null justifying real discrete spectra.}
		\end{figure} 
One can readily check that if $x=a+ib$ is a root of $V_{BB}(x,N)=E$, thus, $-(ai-b)^N=E$. The complex conjugate of this equation would be $-(-ai-b)^N=E$ and hence, $-a+ib$ is the other turning point of this  complex PT-symmetric pair.
Here  all of $E, N, a, b$ are real. The most interesting feature of $V_{BB}$ is that when the parameter $N$ increases there are more than one roots of the equation $-(ix)^N=E$, which are given as
\begin{equation}
x_k=-iE^{1/N}y_k, \quad y_k= e^{i(2k+1)\pi/N},\quad k=0,1,2...
\end{equation}
setting the energy dependence apart, here $y_k(N)$ are the effective turning points. Since, $N$ is not essentially an integer, one is advised to cross check whether every considered $x_k$ really satisfies $-(ix_k)^N=E$. Also, as discussed above, if $x_k$ is a turning point so is $-x^*_k$. For a fixed value of $N$, we wish to define the maximal turning point (MxTP) as the one root $x_k$, which has the absolute value of the real part maximum. We find that for the  parametric regimes
[2,4), (4,8 ] and (8,12] of $N$, $k$ is 0, 1 and 2, respectively. In this regard the PT-symmetric  pair of turning points corresponding to $N\in[2,4)$ being unique is  also maximal. Finally, the pair of turning points to be used is $(-x^*_k,x_k)$.

In Fig. 3,  we show real and imaginary parts of $x_k$ as a function of $N$ by setting $E=1$. The solid (real part) and dashed (imaginary part) lines indicate the variation of the Minimal Turning Points (MnTP) as a function of $N$, which respectively coincides with MxTP:  dots (real part) and triangles (imaginary part) only for $N \in [2,4].$ We would like to remark that $E_n^{BB}$ (1) arises due to the MnTP (2) which may be readily  checked to be the same as the MxTP for $N\in [2,4].$ For, $N \ge 4$, the analytically continued $E_n(N)$ which are proposed in [1,3-7,9-12] can be seen to be arising from semi-classical quantization using MnTPs: analytically continued real (solid) and imaginary(dashed) parts.  

Another interesting feature of MxTP is that their imaginary part is the least so MxTP lie close to real line for a fixed value of $E$ and $N$. Being closer to the real line their contribution to semi-classical eigenvalues is most dominant. This is evident from our Fig. 4 and Table I, where we compare the semi-classical eigenvalues (M1) with those obtained by Dirichlet boundary condition (M2/M3).

\begin{figure}[t]
\centering
\includegraphics[width=17 cm,height=10.cm]{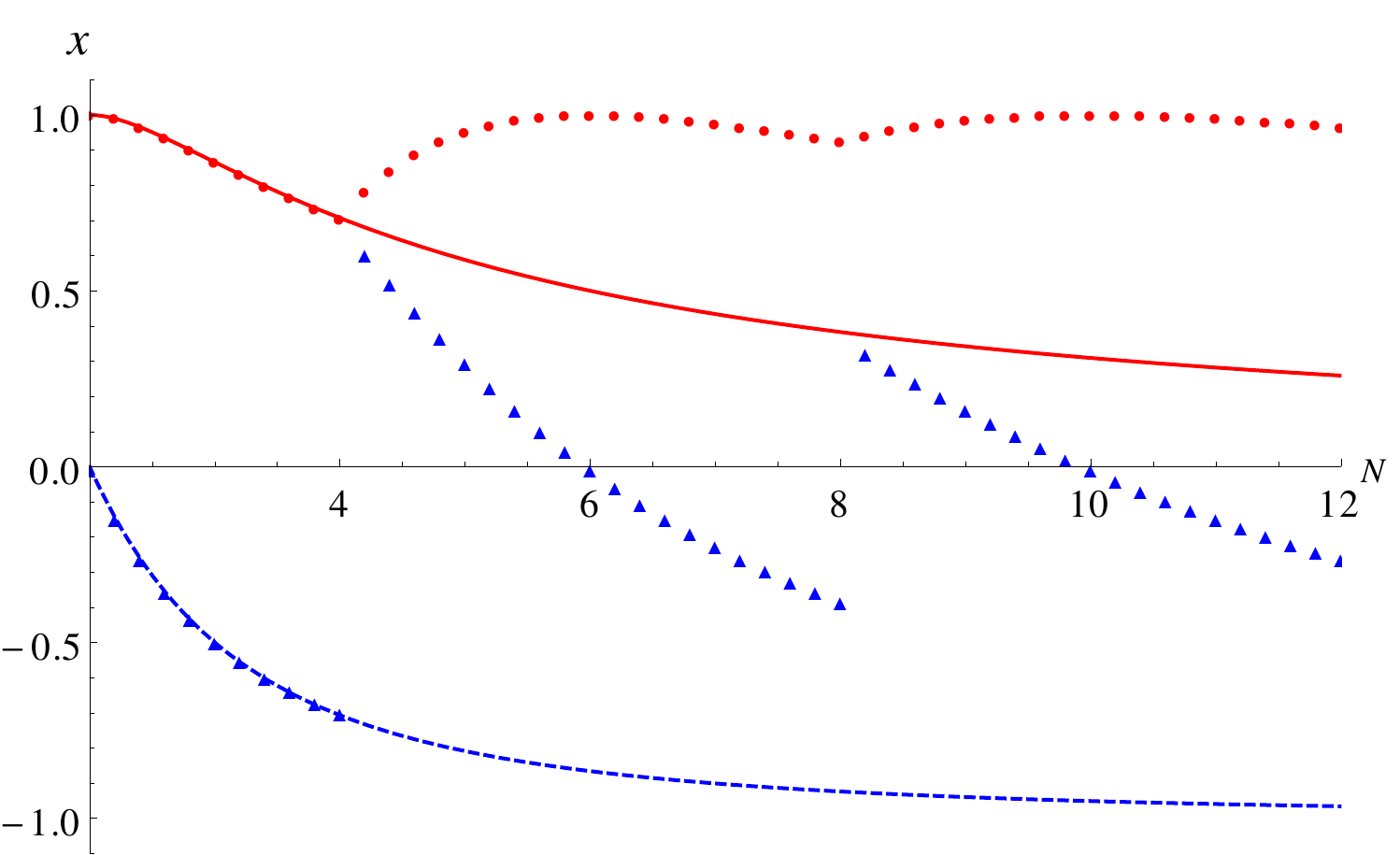}
\caption{The classical complex  turning points x for $V_{BB}(x,N)$ for $E=1$ as a function of $N$. The solid (real part) and dashed (imaginary part) curves denote the minimal turning points (MnTP) considered earlier [1] (see Eq. 2).  Dots (real part) and triangles (imaginary part) denote the maximal turning points (MxTP) proposed here (see Eq. (4)).  Notice the coincidence of the two for $2\le N \le 4$. Also, notice the closeness of triangles to the real line. Whereas, the dashed line is away from the real axis.} 
\end{figure} 
\section{Three Methods: M1, M2 and M3}
In this paper, we wish to compare the real discrete spectra of $V_{BB}(x,N)$
obtained by four methods. The first one we denote by M0 which is represented by the formula (1) as proposed in [1] using the complex turning points as shown by solid (real part) and dashed lines (imaginary part) in Fig. 3. M1 is due to  CWKB but using the  MxTP (see dots and triangles in Fig. 3) proposed above. Methods M2 and M3 are  employing numerical integration of Schr{\"o}dinger equation with DBC (Dirichlet Boundary Condition) and the matrix diagonalization in harmonic oscillator basis, respectively. In the following, we   discuss M1, M2 and M3.
\subsection{M1: Complex WKB using Maximal Turning Points} We proceed to find the general formula for  $E_n^{\mbox{MxTP}}$ for $V_{BB}(x,N)$ arising from the MxTP proposed above (4). The semi-classical action
integral, $I$, in CWKB2
\begin{equation}
I=\int_{-x^*_k}^{x_k} \sqrt{E+(ix)^N} dx = \pi (n+1/2),
\end{equation}
is to be transformed using $x=-iE^{1/N}y$, it is then we split $I$ into two parts as
\begin{equation}
I =-iE^{1/2+1/N} \int_{y_k^*}^{y_k} \sqrt{1+y^N} dy= -i E^{1/2+1/N}\left[\int_{y_k^*}^{0} \sqrt{1+y^N} dy+\int_{0}^{y_k} \sqrt{1+y^N} dy\right]. 
\end{equation}

\begin{figure}[t]
\centering
\includegraphics[width=17 cm,height=10cm]{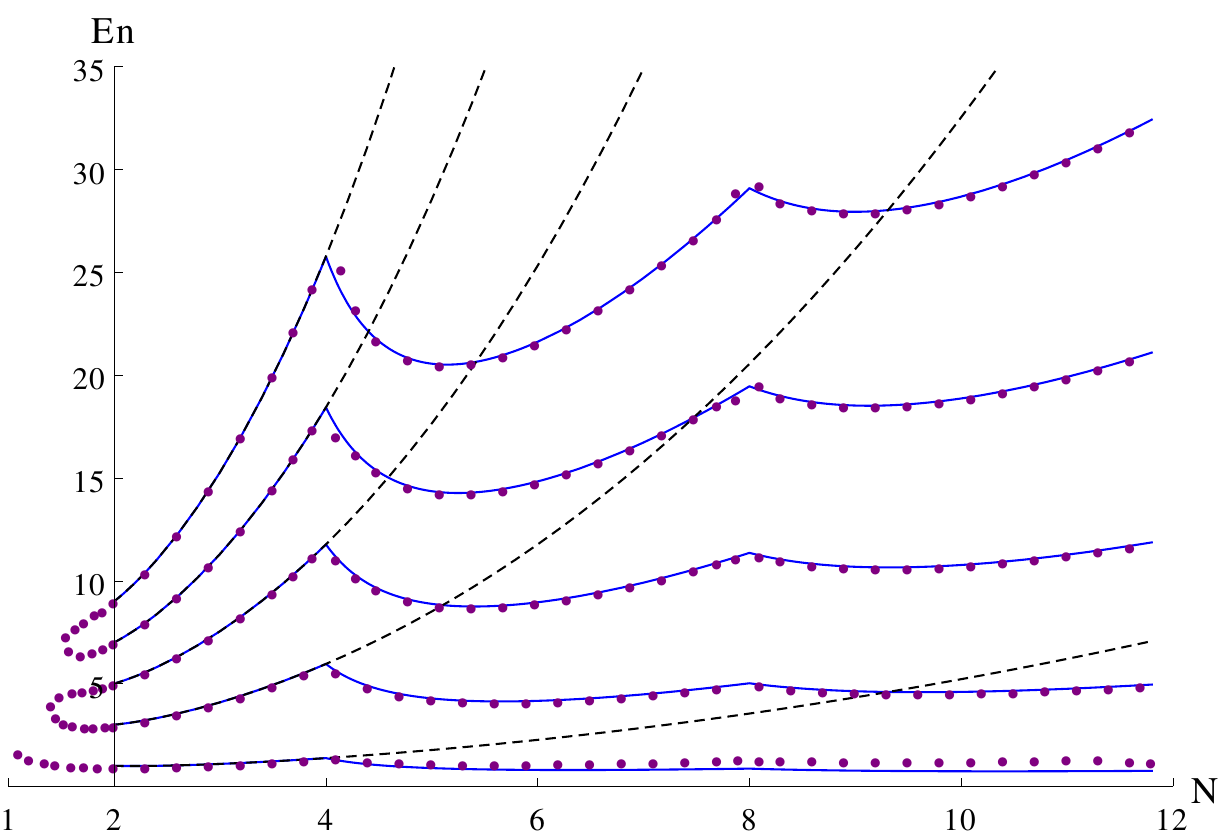}
\caption{First five real discrete eigenvalues of $V_{BB}(x, 1 < N<12)$:
Dashed curves are from  M0 using Eq. (1) [1], solid curves are due to M1 (8) and dots represent the methods M2 and M3, as these two results almost coincide (see Table I). Dots and solid curves show a good agreement for $2\le N<12$. Not shown here are the erratic (non-convergent) results due to $M2$ and $M3$, when we get  closer to $N=4$ and $N=8$, as one does not expect the Hermitian potential barriers $V(x)=-x^4, -x^8$ to possess Dirichlet spectrum. When $N=6, 10$, $V_{BB}$ are Hermitian potential wells we recover the expected (13) [17] eigenvalues. Notice that the  dots, solid and dashed curves coincide  in [2,4) and at $N=4$ solid and dashed curves cross each other to  deviate largely there after.} 
\end{figure}

Next using $y=sy_k^*$ and $y=sy_k$ in the first and second integral, respectively, we get
\begin{equation}
\pi (n+1/2)=-iE^{1/2+1/N}(y_k-y_k^*)\int_{0}^{1} \sqrt{1-s^N} ds= 2 E^{1/2+1/N} \Delta^{\mbox{MxTP}}(N)  \int_{0}^{1} \sqrt{1-s^N} ds,
\end{equation}
where we define $\Delta^{\mbox{MxTP}}(N)=-i(y_k-y^*_k)/2$ (4).  We finally get
\begin{equation}
E^{\mbox{MxTP}}_n(N) = \left[ \frac{\Gamma(3/2+1/N) \sqrt{\pi}(n+1/2)}{\Delta^{\mbox{MxTP}}(N)\Gamma(1+1/N)}\right ]^{\frac{2N}{N+2}},~ n=0,1,2,..,\\
\end{equation} where
\[
\Delta^{\mbox{MxTP}}(N)=\begin{cases}
              \sin (\pi/N)   ~~~~~~          (2 \le N \le 4) \\
               \sin (3\pi/N) ~~~~~           (4 < N \le 8)\\
               \sin (5\pi/N) ~~~~~            (8 < N \le 12),
            \end{cases}
\]
is a continuous function of $N$. Very interestingly, at $N=4,8$, $E_n(N)$ becomes non-analytic as the first derivative is discontinuous.
We show the variation of real and imaginary parts of $x(N)$ (1) in Fig. 3 for $E=1$. Notice that MnTP (2) and MxTP (4) coincide only for $N\in[2,4]$. The spectra of $V_{BB}(x,N)$ arising from (1) (dashed curves) and (8) (solid curves) are shown in Fig. 4, these thus coincide for $N \in [2,4].$ 
{\subsection{M2: Numerical integration of Schr{\"o}dinger equation with Dirichlet Boundary Condition}
The next crucial question is whether the eigenspectra due to the new formula (8) will be consistent with the exact eigenvalues obtained by solving Schr{\"o}dinger equation by imposing the most conventional Dirichlet boundary condition: $\psi(\pm \infty)=0.$ We need to show that the evolution of eigenvalues $E_n(N)$ for
\begin{equation}
\frac{d^2\psi}{dx^2}+[E+(ix)^N]\psi(x)=0, \quad \psi(\pm \infty)=0,
 \end{equation}
agrees with our semi-classical formula for $E^{\mbox{MxTP}}_n(N)$ (8).

We seek $\psi(E,x)=A u(E,x)+ B v(E,x)$ for $-d<x<d$ and $\psi(E,\pm d)=0$, where $d$ is the chosen asymptotic distance. The functions $u(E,x)$ and $v(E,x)$ are two linearly independent solutions of (9). We start numerical integration from $x=0$  up to $x=\pm d$ on both sides taking $u(E,0)=1, u'(E,0)=0$ and $v(E,0)=0, v'(E,0)=1$. We finally get the eigenvalue equations as 
\begin{equation}
u(E,d)v(E,-d)-u(E,-d)v(E,d)=0 \quad {\mbox or} \quad \frac{u(E,d)}{v(E,d)}=\frac{u(E,-d)}{v(E,-d)}.
\end{equation}
These two equations are indeed equivalent yet one may be more convenient than other in numerical computations for various values of $N$. To be sure, we use both for the correctness of a result. Another point in these calculations is a proper choice of $d$, such that eigenvalues do not change appreciably as we change the value of $d$ for a given accuracy. For an accuracy of $10^{-2}$, we find that $d= 10$ is an optimum value, we check that our results do not change at least up to two decimal places as we take $d=$ 9 to 11. Going for a better accuracy adds only to computational time
and no other complication. Also since we are interested in commenting on the correct trend of the
parametric evolution of $E_n(N)$ ( analytically continued (dashed) curves  versus the solid curves in Fig. 4), a better accuracy is not indeed a concern here.
\subsection{M3: Matrix Diagonalization of $H=p^2/2\mu+V_{BB}(x,N)$ in harmonic Oscillator basis}
Choose $2\mu=1=\hbar^2, \omega=2$, such that $H_{HO}=-\frac{d^2}{dx^2}+x^2$ and its eigenvalues are $(2n+1)$ with eigenstates as $\ket{n}=(2^n n! \sqrt{\pi})^{-1/2}~ e^{-x^2/2} H_n(x)$. We  write the matrix element $ix_{m,n}$ , $p_{m,n}$ in this  basis  as :
\begin{equation}
X=ix_{m,n}=\frac{i}{\sqrt{2}}[\sqrt{n+1} ~ \delta_{m,n+1}+\sqrt{n} ~ \delta_{m,n-1}],\quad
p=p_{m,n}=\frac{i}{\sqrt{2}}[\sqrt{n+1} ~\delta_{m,n+1}-\sqrt{n} ~\delta_{m,n-1}].
\end{equation}
We have calculated  matrix elements   $\mel{m}{-(ix)^N}{n}$ in this basis to construct the Hamiltonian matrix. We have found ``MatrixPower" (in ``Matlab") 2 and $N$ of $p$ and $X$, respectively to diagonalize the matrix :
\begin{equation}
H=p^2-X^N.
\end{equation}
We diagonalize the matrix $H_{{\cal N} \times {\cal N}}$ (12)  using ${\cal N}$ as $\sim 1500$, we ensure  an accuracy better than $10^{-2}$ for all the eigenvalues given in the Table I. Better accuracy requires higher values of ${\cal N}$, more computational time no other complication. 
\section{Results and Discussion}
Our results on the Dirichlet spectra of $V_{BB}(x, 2\le N<12)$ due to the method of numerical integration (M2) and matrix diagonalization (M3) may not be very accurate as we have achieved an accuracy of $10^{-2}$ or more. With this limitation, in most cases, in the Table I, the agreement between these two set of eigenvalues even up to third or fourth places of decimal is satisfying. The ground state eigenvalues $E_0(N)$ are underestimated more by our exact CWKB formula (8). This is a usual feature of semi-classical methods. Moreover, the overall good agreement between dots (M2/M3) and solid curves (M1) in Fig. 4, is the testimony to consistency of the Dirichlet spectra presented in Fig. 4. We would like to remark that this consistency is thought provoking as $E_n(N)$ appears to have been obtained by piecing together three parts (8) and consequently the $E_n(N)$ are continuous but non-differentiable at two IPs: $N=4,8$. The coincidence of all four spectra up to $N=4$ in Fig. 4 is re-assuring. 

For $N=4$ and $N=8$, $V_{BB}$ becomes a flat-top Hermitian potential barrier [5, 19] devoid of Dirichlet spectrum. So, when we get very close to these $N$ values, the eigenvalues are non convergent. This shows a fuzzy  dependence on the choice of the asymptotic distance $d$ in the numerical integration method (M2) and also on the size ($\cal N$) of the matrices in the method of diagonalization (M3). However, for $N=3.8, 4.2;$ see the Table I and Fig. 4, we get eigenvalues which converge well. Interestingly,  before the advent of complex PT-symmetric quantum mechanics in the year 1995 [18], a semi-classical quantization identical to Eq. (5) has been proposed to find the semi-classical discrete spectrum of perfect transmission (zero reflection) energies for the Hermitian potential barriers, e.g $V(x)=V_0(1+x^4)^{-1}$ which has two pairs complex turning points $(-z^*, z)$. Further, in the light of the discussion in Refs. [5, 19], we conjecture that our formula (8) for $N=4, 8$, where we choose only MxTP out of three pairs of complex turning points gives us the discrete spectrum of reflectivity zeros. Due to this very reason, see in Fig. 4,  we do not get the Dirichlet spectrum for  $N=4, 8$ and for the values of $N$ very close to these values, when we use  methods M2/M3. 

\begin{table}[t]
	\caption{\label{tab:table1} First five real discrete eigenvalues of $V_{BB}(x,N)$ for various values of $N$ due to methods M1, M2, M3. As discussed in the text below (13), M1 (8) also represents the well known eigenvalues for $N = 6, 10 (13)$ [17]. For $E_0$, the deviation of M1 values with those of M2/M3 is attributed to semi-classical approximation (M1).}
	\begin{tabular}{cccccccccc}
		\hline
		\begin{tabular}{@{}c@{}} $M1$ \\ $M2$ \\ $M3$ \end{tabular}  & $N=2.5$ & $N=3$ & $N=3.8$ & $N=4.2$ & $N=5.3$  & $N=6$\footnote{$^{,b}$ Notice, the important recovery of eigenvalues (13) [17] in these two Hermitian cases for $N = 6,10$. The analytically continued spectra [1,3-7,9-12], miss out on these eigenvalues.} & $N=6.8$ & $N=10^b$ \\
		
		\hline
		& 1.0112    &  1.0942 & 1.3102 & 1.1983 & 0.8459 & 0.8008 & 0.8041 & 0.7365 \\
		\begin{tabular}{@{}c@{}} $E_0$ \end{tabular}  & 1.0490    &  1.1563 & 1.4035 & 1.3640 & 1.1427 & 1.1448 & 1.2035 & 1.2988  \\
		& 1.0489    &  1.1562 & 1.4035 & 1.3639 & 1.1427 & 1.1448 & 1.1951 & 1.2986 \\
		
		\hline
		&3.4275  & 4.0895 & 5.5276  & 5.3086 & 4.1702  & 4.1612 & 4.3922 & 4.5960 \\
		\begin{tabular}{@{}c@{}} $E_1$ \end{tabular}  &3.4345  & 4.1092  & 5.5694 & 5.3024  & 4.2875 & 4.3386 & 4.6245 & 5.0979 \\
		&3.4345  & 4.1092  & 5.5687 & 5.3024  & 4.2875 & 4.3386 & 4.6245 & 5.1024\\
		
		\hline
		&6.0461 & 7.5490 & 10.7954 & 10.6061 & 8.7558 & 8.9535 & 9.6726 & 10.7678 \\
		\begin{tabular}{@{}c@{}} $E_2$ \end{tabular}  & 6.0517 & 7.5623 & 10.8244 & 10.8051 & 8.8422 & 9.0731 & 9.8301 &  11.1543 \\
		& 6.0517 & 7.5629 & 10.8247 & 10.8046 & 8.8422 & 9.0731  & 9.8301  & 11.1539  \\
		
		\hline
		& 8.7869 & 11.3043 & 16.7771 & 16.7315 & 14.2720 & 14.8316 & 16.2696 & 18.8657 \\
		\begin{tabular}{@{}c@{}} $E_3$ \end{tabular}  & 8.7910 & 11.3144 & 16.7995 & 16.2691 & 14.3480 & 14.9352 & 16.4083 & 19.1889 \\
		& 8.7907 & 11.3108 & 16.7991 & 16.3962 & 14.3480 & 14.9352 & 16.4083 & 19.1884 \\
		
		\hline
		& 11.6175 & 15.2833 & 23.3203 & 23.5184 & 20.5574 & 21.6224 & 23.9914 & 28.6801 \\
		\begin{tabular}{@{}c@{}} $E_4$ \end{tabular} & 11.6207 & 15.2916 & 23.3380 & 23.9034 & 20.6217 & 21.7142 & 24.1155 & 28.9715\\
		& 11.6206 & 15.2960 & 23.3113 & 23.9034 & 20.6217 & 21.7142 & 24.1154 & 28.9715 \\
		\hline
		\label{tab:dis_size}   
	\end{tabular}
\end{table}

A fair  agreement (see Fig. 4 and Table I)  of the  eigenvalues from  our methods M1,M2 and M3 is one of the most striking features of the present work. For the Hermitian potentials  $V_H(x,N)= |x|^N$ the simple WKB eigenvalues are well known as [17].
\begin{equation}
E_n^{H}(N) = \left[ \frac{\Gamma(3/2+1/N) \sqrt{\pi}(n+1/2)}{\Gamma(1+1/N)}\right ]^{\frac{2N}{N+2}}, \quad E_n^{H}(\infty)\sim \frac{(n+1/2)^2\pi^2}{4},~ n=0,1,2,...
\end{equation}
 Notice a small slip in Eq. 5 of Ref.[1]: $(n+1)^2$ for $(n+1/2)^2$. It needs to be remarked here that for these symmetric Hermitian potentials for $N>2$ too have complex pairs of classical turning points wherein the used [14] real pair: $x=\pm E^{1/\nu}$ are again the maximal turning points. One can readily check that for $V(x)=x^6$ and $x^{10}$, the above equation (13) and our result (8) coincide, whereas Eq.(1) [1] deviates from (13). These deviations can be easily observed in Fig. 4 as the dashed lines leave the solid lines (8) and dots (M2,M3) for $N>4$. Also see the Table I, in this regard. The $\Delta^{\mbox{MxTP}}(N)$ in Eq. (8) can be generalized as $(2K+1)\pi/N $ if $4K < N \le 4K+4, K=1,2,3...$, then for $N=2+4K$, $V_{BB}(x,N)$ is real Hermitian having real discrete spectrum given by (8) or (13).

A CPTSP may have several families (branches) of real discrete spectrum [8]. We suggest that seeking Dirichlet spectrum brings the much required uniqueness in  PT-symmetric quantum mechanics. In more interesting models namely the scattering potential wells [14] and other [15], Dirichlet spectrum may itself have two branches (identified by quasi parity [20]) and by putting them together one observes coalescing of eigenvalues at the exceptional point(s) of the complex potential. In even more interesting models like Scarf II [21] and shifted harmonic oscillator [22], in addition to the coalescing one may observe crossing(s) of eigenvalues in one dimension! however, the corresponding eigenstates are linearly dependent [21] shunning degeneracy in one-dimension. A few  solvable or quasi-exactly solvable potentials: $V(x)=e^{2ix}/2 ~[23], V(x)=-(\xi \cosh 2x-iM)^2  ~[24], V(x)=-(\zeta \sinh 2x-iM)^2$ ~[25] and a complex Coulomb potential [26] are models of CPTSP  having this specialty  that their eigenstates do not vanish  on the real line, instead they vanish on some contour, the results will depend on the choice of the contour.  Additionally, their eigenstates are not $L^2$-integrable and we point out that their [23-26] Dirichlet spectrum is not known so far.

\section{Conclusion}
We would like to conclude that this paper has actually accomplished the long due task of finding quantum mechanically, the  most formidable (Dirichlet) real discrete spectra $E_n(N)$ of the paradigm model of Complex PT-Symmetric Potential (CPTSP) for the parameter $N \in [2,4) \cup (4,8) \cup (8,12)$. Here our results differ from the existing ones for $N>4$, remarkably we reproduce the expected spectrum for the Hermitian wells ($N=6,10$). $N=4,8$ turn out to be Isolated Points (IPs) where the parametric derivative of $E_n(N)$ is discontinuous and the Dirichlet spectrum is null. To the best of our knowledge this paradigm model is the first realization of Kato's IPs. For $N$ values close to 4 and 8, better numerical methods and algorithms need to be devised in this regard. New semi-classical quantization methods when there are more than one pair of complex turning points are highly desirable. However, our proposed concept of the pair of maximal  turning points is thought provoking which has worked very well in producing the real discrete spectrum of the paradigm model of CPTSP. The semi-classical coincidence of Dirichlet spectrum and reflectivity zeros for $V(x)=-x^4, -x^8$ is really intriguing as it is absent in the orthodox methods of finding bound states. Lastly, we would like to re-emphasize that Kato's concept of IPs of non-analyticity in
the continuous parametric evolution of eigenvalues, has provided us at least another way to look at the real discrete spectrum of the paradigm model of CPTSP. This is in contrast to the analytic continuation of eigenvalues for $N>4$ done so far.

\section*{\Large{Acknowledgment}} We thank Prof. Carl M. Bender for his critical remarks on two (previous) versions of this work.

\end{document}